\documentclass[twoside,11pt]{article}
\usepackage{jmlr2e}

\RequirePackage{graphicx}
\usepackage{bbm}
\usepackage{amssymb}
\usepackage{amsmath}
\usepackage{mathabx}
\usepackage{subcaption}
\usepackage{natbib}
\usepackage{algorithm2e}
\usepackage{algpseudocode}
\usepackage{apptools}
\usepackage{cleveref}
\usepackage{mathcomp}

\usepackage{booktabs}
\usepackage{multirow}
\usepackage{longtable}
\usepackage{caption}
\RestyleAlgo{ruled}

\SetKwInput{KwData}{Input}

\ShortHeadings{CIs for Random Forest Permutation Importance with Missing Data}{Föge and Pauly}
\firstpageno{1}

\begin{document}

\title{Confidence Intervals for Random Forest Permutation Importance with Missing Data}

\author{\name Nico Föge \email nico.foege@ovgu.de \\
       \addr Institute for Mathematical Stochastics\\
       Otto-von-Guericke University Magdeburg\\
       Magdeburg, 39106, Germany\\[1ex]
       Research Center Trustworthy Data Science and Security\\
       University Alliance Ruhr, 44227, Germany
       \AND
       \name Markus Pauly \email markus.pauly@tu-dortmund.de \\
       \addr Department of Statistics\\
       TU Dortmund University\\
       Dortmund, 44227, Germany\\[1ex]
       Research Center Trustworthy Data Science and Security\\
       University Alliance Ruhr, 44227, Germany}
       
\editor{}

\maketitle

\begin{abstract}
Random Forests are renowned for their predictive accuracy, but valid inference -- particularly about permutation‐based feature importances -- remains challenging. Existing methods, such as Ishwaran et al.’s (2019) confidence intervals (CIs), are promising but assume complete feature observation. However, real‐world data often contains missing values. In this paper, we investigate how common imputation techniques affect the validity of Random Forest permutation‐importance CIs when data are incomplete. Through an extensive simulation and real-world benchmark study, we compare state‐of‐the‐art imputation methods across various missing‐data mechanisms and missing rates. Our results show that single-imputation strategies lead to low CI coverage. As a remedy, we adapt Rubin’s rule to aggregate feature‐importance estimates and their variances over several imputed datasets and account for imputation uncertainty. Our numerical results indicate that the adjusted CIs achieve better nominal coverage.
\end{abstract}



\newpage
\section{Introduction}
Random Forests, introduced by \citet{Breiman2001}, are powerful tree-based ensemble methods, widely known for their robustness and predictive accuracy across diverse applications. In contrast to  (generalized) linear regression models -- which offer interpretable parameter and variance estimates that support formal statistical inference -- Random Forests rely on feature‐importance measures that lack well‐developed inferential statistics, such as CIs and hypothesis tests. 
The absence of analytical solutions for variance estimation and the intricate dependencies within tree-based predictions, making statistical inference far less straightforward.

Recent advancements have sought to address these challenges by developing rigorous frameworks for inference in Random Forests. 
\citet{Scornet} and \citet{blum2024consistency} proved consistency for Random Forest predictions. Moreover, \citet{wager2014confidence} and 
\citet{mentch2016quantifying} established asymptotic normality and variance estimators for Random Forest predictions, which can be used to construct prediction intervals. However, these advances address the prediction level rather than inference on the feature level. 

But feature importances are crucial for practitioners who want to understand which covariates drive predictions. 
\citet{BurimundMarkus} demonstrated the consistency of Random Forest Permutation Importance (RFPIM), affirming its reliability as a measure of feature significance - though under rather strong assumptions. Based on RFPIMs, \citet{ishwaran2019standard} proposed confidence intervals (CIs) for interpretable population feature importance scores. 
The approach is implemented in the \textbf{R} package $\texttt{randomForestSRC}$,
that has been downloaded more than 900.000 times (18th July 2025).
Although these numbers and the simulation results on the validity of these RFPIM-based CIs presented in \citet{ishwaran2019standard} are promising, the approach 
lacks a formal proof of the implied asymptotic normality of RFPIMs. 
The recent preprint \citet{foge2024central} made a first theoretical contribution to close this gap by proving a central limit theorem for RFPIMs. 
However, the results rely on strong conditions, which may limit their applicability in practice. Together, these developments highlight a growing body of research aimed at bridging the gap between the predictive power of Random Forests and the need for interpretable statistical inference. 

Despite these promising developments, existing inference procedures assume that the features are fully observed. 
However, real-world datasets are rarely complete, and missing data pose a challenge to the applicability of these inferential methods. 
 The performance of imputation methods in the context of Random Forest-based inference, particularly for Ishwaran’s CIs for population feature importance scores, remains underexplored. This gap raises two research questions:  (i) How do existing imputation methods affect the coverage and validity of Ishwaran’s feature‐importance CIs when data are missing? (ii) Can one devise an imputation‐inference strategy that preserves nominal coverage?
 To address these questions, we conduct an extensive simulation study to 
 compare different state-of-the-art imputation techniques -- such as \texttt{mice} \citep{vanBuuren2011mice}, missforest \citep{stekhoven2012using}, and \texttt{mixgb} an XGBoost‐based imputation \citep{deng2024xgboost} -- and evaluate their impact and the robustness of Ishwaran’s CIs under various missing data scenarios.
 Our results show that simply imputing missing values by one of those approaches substantially reduce the CI coverage. 
 
 To tackle this, we propose an alternative inference approach based on Rubin’s rule \cite{Rubin1987,rubin2018multiple} for combining estimates across multiple imputed datasets: By treating each imputed dataset as a separate realization and aggregating the resulting importance estimates and variances, we derive adjusted CIs that better account for imputation uncertainty. Through extensive simulation, we show that this approach leads to improved coverage rates, suggesting that combining multiple imputation with 
 Rubin’s rule is essential for valid RFPIM-based CIs in the presence of missing data.
 
The rest of the paper is structured as follows.
In \cref{Section 2} we formally introduce the RFPIM and the CIs. \cref{Section 3} presents the investigated imputation methods, including their adaptation with Rubin's rule.
\cref{Section 4} describes the settings of our simulation study and its results are presented in \cref{Section 5}.
To also investigate the performance on real-world data we conduct a benchmark study in \cref{Section 6}.
The paper closes with a discussion (\cref{Section 7}).

\section{Permutation Importance and Ishwarans Confidence Intervals}
\label{Section 2}
We consider a regression setting with a response variable \( Y \in \mathbb{R} \) and a feature vector \( \mathbf{X} = (X_1, \ldots, X_p) \in \mathbb{R}^p \). The underlying regression function is denoted by \( m(\mathbf{x}) = \mathbb{E}[Y | \mathbf{X} = \mathbf{x}] \).
A Random Forest consists of M tree predictors, each generated using random mechanisms which we describe by random vectors \( \mathbf{\Theta}_1, \ldots, \mathbf{\Theta}_M \). We write 
\( m_{n,1}(\mathbf{x}, \mathbf{\Theta}_t, \mathcal{D}_n) \) for the prediction of tree  \( t \) on an input covariate vector \( \mathbf{x} \in \mathbb{R}^p \). Here \( \mathcal{D}_n \) is the training dataset consisting of \( n \) i.i.d. copies \( (\mathbf{X}_1, Y_1), \ldots, (\mathbf{X}_n, Y_n) \) of \( (\mathbf{X}, Y) \). The \textit{Random Forest prediction} on \( \mathbf{x} \) is then defined as the ensemble prediction
\[
m_{n,M}(\mathbf{x}, \mathbf{\Theta}_1, \ldots, \mathbf{\Theta}_M, \mathcal{D}_n) = \frac{1}{M} \sum_{t=1}^M m_{n,1}(\mathbf{x}, \mathbf{\Theta}_t, \mathcal{D}_n).
\]
Each tree is trained on a bootstrap subsample drawn from \( \mathcal{D}_n \). The excluded points form the Out-of-Bag (OOB) set \( \mathcal{D}_n^{-(t)}(\mathbf{\Theta}_t) \) of tree \(t\).

The permutation importance for the \( j \)-th feature, \( j \in \{1, \ldots, p\} \), quantifies the contribution of \( X^{(j)} \) to the predictive accuracy of the Random Forest. To compute it, the values of the \( j \)-th feature in the OOB sample \( \mathcal{D}_n^{-(t)}(\mathbf{\Theta}_t) \) are randomly permuted. To introduce it more formally, we construct a permuted feature vector for each observation \( (\mathbf{X}_i, Y_i)\) with \( i \in \mathcal{D}_n^{-(t)}(\mathbf{\Theta}_t) \):  \[ \widetilde{\mathbf{X}}_i^{j} = \left(X_i^{(1)}, \ldots, X_i^{(j-1)}, \widetilde{X}_i^{(j)}, X_i^{(j+1)}, \ldots, X_i^{(p)}\right). \] 
Here, \( \widetilde{X}_i^{(j)} \) is a randomly selected value from the set of \( j \)-th feature values in the OOB sample. \cite{ishwaran2019standard} define the \textit{Tree Permutation Importance} for the $t$-th tree and a loss-function $\ell$ as

\begin{align}
\label{tree importance}
    I(j,\mathbf{\Theta_t},\mathcal{D}_n)
    =
    \frac{1}{\gamma_n}\sum_{i\in \mathcal{D}_n^{-(t)}(\mathbf{\Theta}_t) }
    \left\{
        \ell \left(Y_i,m_{n,1}\left(\widetilde{\mathbf{X}}_i^j,\mathbf{\Theta_t},\mathcal{D}_n\right)\right)
        -
           \ell \left(Y_i,m_{n,1}\left(\mathbf{X}_i,\mathbf{\Theta_t},\mathcal{D}_n\right)\right)
    \right\},
\end{align}

\noindent where $\gamma_n=\left\vert \mathcal{D}_n^{-(t)}(\mathbf{\Theta}_t)\right\vert$ 
is the size of the OOB set for tree $t$. Since we
focus on a regression setting, $\ell$ is chosen as the squared loss $\ell(x,y)=(x-y)^2$. Averaging (\ref{tree importance}) across all trees $t=1,\dots,M$ yields 
the \textit{Random Forest Permutation Importance Measure (RFPIM)}
\begin{align}\label{eq: RFPIM}
    I_{n,M}^{OOB}(j,\mathcal{D}_n)=I_{n,M}^{OOB}(j,\mathbf{\Theta}_1,\ldots,\mathbf{\Theta}_M,\mathcal{D}_n)
    =
    \frac{1}{M}\sum_{t=1}^M
       I(j,\mathbf{\Theta_t},\mathcal{D}_n).
\end{align}
Due to the dependence structure of the OOB-samples, its rather complex to estimate the variance of \eqref{eq: RFPIM}. This makes it hard to construct CIs
for the true population \textit{Feature Importance (Scores)}
\begin{align*}
    I(j):=
    \mathbb{E}\left[\Big(Y_1- m \left(\mathbf{X}_{j,1}\right)\Big)^2\right]
    -
    \mathbb{E}\left[\Big(Y_1- m \left(\mathbf{X}_1\right)\Big)^2\right].
\end{align*}
Here, $\mathbf{X}_{j,1}=\left[X_{1}^{(1)},\ldots,X_{1}^{(j-1)},X^{(j)},X_{1}^{(j+1)},
\ldots,X_1^{(p)}\right]$ and $X^{(j)}$ is an independent copy of $X_1^{(j)}$ \citep{gregorutti2017correlation}.
\cite{ishwaran2019standard} compared several variance estimators and recommend the delete-$d$ Jackknife estimator, which has been implemented in the $\texttt{RandomForestSRC}$ \textbf{R} package \citep{ishwaran2007random, ishwaran2008random,ishwaran2025fast,R}. The delete-$d$ Jackknife is computed as in
\cite{ishwaran2019standard} and the corresponding Algorithm~\ref{alg:example} is displayed below.\\

\begin{algorithm}[H] 
\SetAlgoLined 
\KwIn{data set $\mathcal{D}_n$, number of repetitions $K$, subsample size $b$ (so delete $d=n-b$), trained Random Forest generated by $\mathbf{\Theta}_1,
\ldots, \mathbf{\Theta}_M$,}
\KwOut{The delete-$d$ variance estimator $v^2_{d,K}$} 
Use the entire dataset $\mathcal{D}_n$ to compute $ I_{n,M}^{OOB}(j,\mathcal{D}_n)$\\
\For{$k \gets 1$ \KwTo $K$}{
    Draw a subsample $s_k$ of size $b$ from $\{1,\ldots, n\}$\;
    Denote the subsample of $\mathcal{D}_n$ containing only the indices of $s_k$ by $\mathcal{D}_{b,s_k} =\{(\mathbf{X}_i,Y_i): i\in s_k\}$\;
    Calculate and save RFPIM $I_{n,M}^{OOB}(j,\mathcal{D}_{b,s_k})$
    using only $\mathcal{D}_{b,s_k}$\;
}
\Return $v_{d,K}^2=\frac{b}{(n-b)K}\sum\limits_{k=1}^K\left(I_{n,M}^{OOB}(j,\mathcal{D}_{b,s_k})- I_{n,M}^{OOB}(j,\mathcal{D}_n)\right)^2$
\caption{Delete-$d$ Jackknife estimator ($d=n-b$) for $\mathbb{V}\text{ar}(I_{n,M}^{OOB}(j))$}
\label{alg:example}
\end{algorithm}

\vspace{\baselineskip}
\noindent 
Using the variance estimator $v_{d,K}^2$,  \cite{ishwaran2019standard} propose the $(1-\alpha)$ CI
\begin{align*}
    \left[I_{n,M}^{OOB}(j,\mathcal{D}_n)-z_{1-\alpha/2}\sqrt{v_{d,K}^2},
    I_{n,M}^{OOB}(j,\mathcal{D}_n)+z_{1-\alpha/2}\sqrt{v_{d,K}^2}\right].
\end{align*}
Here, $z_{1-\alpha/2}$ is the $(1-\frac{\alpha}{2})$-quantile of the standard normal distribution. This normal approximation was justified by promising simulations \citep{ishwaran2019standard}. In \cite{foge2024central} a theoretical justification - though under restrictive assumptions on the hyperparameters - is also given. 
In summary, this approach constructs CIs via RFPIM and the delete-$d$ Jackknife variance estimator, under the assumption that all features are fully observed. We next investigate the challenge of missing data, and how imputation before training may impact the validity of these CIs.

\section{Missing Data and Imputation Approaches}
\label{Section 3}
Missing data are ubiquitous in empirical research across disciplines. 
Discarding incomplete observations in the analysis (complete‐case or listwise deletion) can introduce bias estimates/predictions or lead to a substantial 
loss of information \citep{schafer1997analysis,little2019statistical}. As a remedy, (multiple) imputation is recommended, particularly when data is missing at random (MAR) or missing completely at random (MCAR) \citep{pepinsky2018note}. In multiple imputation (MI), missing values are replaced by draws from the predictive (imputation) distribution multiple times, generating $R$ complete datasets. Final estimates and standard errors are then combined across these datasets using Rubin’s rule to reflect uncertainty from both sampling and imputation \citep{Rubin1987,rubin2018multiple}.

In general, a variety of imputation techniques exist, ranging from simple single‐imputation approaches (e.g., mean imputation or regression imputation) to more advanced techniques such as multiple imputation by chained equations (MICE) \citep{vanBuuren2011mice,white2011multiple}. The choice of method depends on the study design, the assumed missingness mechanism, the scale of the features, and the primary inferential goals. For comprehensive overviews and statistical foundations, we refer to \citet{Rubin1987,schafer1997analysis,vanBuuren2011mice,white2011multiple,rubin2018multiple,
little2019statistical,thurow2021imputing,buczak2023analyzing}. 

For our selection of imputation methods, we deliberately focus on modern, state‐of‐the‐art approaches that have demonstrated favorable properties  in a wide range of scenarios and are available in
 robust and well-maintained \textbf{R} packages. Moreover, as we particularly investigate the effect of both single and multiple imputation, we wanted complementary imputation techniques offering both imputation variants, or at least allow 
  for predictive‐mean‐matching (PMM).
   PMM ensures that only observed values are used for imputation, thus preserving the original data distribution and avoiding implausible values \citep{rubin1987multiple,little1988missing,white2011multiple}. Moreover, PMM is less sensitive to model misspecification compared to purely parametric approaches and is well-suited for non-normally distributed variables \citep{morris2014tuning}.
 We thus selected three imputation that fulfill these requireements:

\textbf{I. Multiple Imputation by Chained Equations (MICE).} MICE iteratively models each feature with missing values using regression based on other features to produce imputations \citep{vanBuuren2011mice,white2011multiple}.
Imputation can be performed using single imputation with Predictive Mean Matching (PMM) or multiple imputation, where multiple plausible values are generated to account for uncertainty. 
MICE with PMM exhibits favorable distributional imputation accuracy \cite{thurow2021imputing,thurow2024assessing} and is still the gold standard imputation method in many quantitative sciences
\citep{schwerter2024evaluatingtreebasedimputationmethods}.

The implementation of MICE is done with the R package \texttt{mice} \citep{vanBuuren2011mice} using the options
$\texttt{m=5,~maxit=5}$ and \texttt{method="pmm"}.

\textbf{II. Chained Random Forests (missRanger).} 
This approach leverages the Random Forest algorithm to sequentially predict each feature with missing values using out-of-bag estimates from Random Forests trained on the other features \citep{stekhoven2012using}. Though it doesn't support multiple imputation, it can deal with mixed-type data and its extension by \citep{Missranger} also supports PMM: after obtaining a predicted value for a missing observation, missRanger samples from a small set of observed values whose out-of-bag predictions are closest. Empirical studies \citep{Stekhoven, waljee2013comparison,BurimundMarkus2,buczak2023analyzing} have shown that missRanger often outperforms parametric-imputation approaches with respect to imputation accuracy and other predictive tasks. However, when applied without PMM, it was also seen to inflate type-I-errors of statistical tests in linear models \cite{ramosaj2020cautionary,schwerter2024evaluatingtreebasedimputationmethods}. 
For the chained Random Forest Imputation we use the \texttt{missRanger} package \citep{Missranger}
which is based upon the fast \texttt{ranger} implementation of Random Forests. We chose the options \texttt{num.trees=200,
~pmm.k=5} and the default values for the rest of the hyperparameters.

\textbf{III. XGBoost-based Imputation (mixgb).}
This algorithm uses the XGBoost algorithm, a gradient boosting method, to estimate missing values within a multiple imputation framework \citep{deng2024xgboost}. Implemented in the \texttt{mixgb} package, mixgb captures complex relationships efficiently, using Predictive Mean Matching (PMM) and subsampling  to enhance imputation accuracy and variability \citep{mixgb}. Multiple imputation generates several plausible datasets  to account for uncertainty \citep{deng2024xgboost}.
 
These are also the same imputation algorithms as considered by \citet{schwerter2024evaluatingtreebasedimputationmethods} who also evaluate inferential properties (estimation bias, type-I-error and power of tests), but in traditional linear‐models. Our goal is to assess how these methods perform when paired with nonparametric Random Forest–based CIs that should better deal with  nonlinear data structures. By including both single- and multiple-imputation variants of \texttt{mice} and \texttt{mixgb}, we can directly compare:

\textbf{(A) Single-Imputations as Quick Fixes} that fill in missing values once 
(with PMM) 
and proceed with the construction of Ishwaran's CIs as if the data were fully observed. In parametric linear models, single imputation often underestimates uncertainty, which can lead to too narrow CIs for parameters \citep{vanBuuren2011mice,white2011multiple}. As we will see later, this approach also causes serious problems for Random Forest-based CIs.

\textbf{(B) More complex Multiple-Imputations} that generate $R$ complete datasets via repeated draws from each imputation model (applying PMM on each draw) and then pool estimates via Rubin’s rule \citep{Rubin1987}. In parametric linear models, this approach leads to valid inference procedures when data is MAR. However, there are neither theoretical reasons nor numerical studies that support this for Random Forest-based CIs. We will close this gap in the present paper.

To explain the heuristic multiple imputation approach based on Rubin's rule for combining estimators \citep{rubin2018multiple} recall that 
\texttt{mice} and \texttt{mixgb} are multiple Imputation methods. Thus, instead of a singe complete dataset, they can generate $R\in\mathbb{N}$ complete datasets. On each of those 
we run the Random Forest and compute the delete-$d$ Jackknife estimator, denoted by
$v_{d,K,r}^2$ ($r=1,\ldots,R$), and the Permutation Importance, denoted by $I_{n,M}^{OOB}(j,\mathcal{D}_n^r)$, where $\mathcal{D}_n^r$ is  the $r$-th
complete dataset. Following Rubin's rule the estimations can be combined by averaging
\begin{align*}
    \widebar{I_{n,M}^{OOB}}(j)=\frac{1}{R}\sum_{r=1}^RI_{n,M}^{OOB}(j,\mathcal{D}_n^r).
\end{align*}
For the combined variance estimation, we have to consider the within-imputation variance
and the between-imputation variance. The within-imputation variance can also be estimated by
averaging
\begin{align*}
   \widebar{v}_{d,K}^2=\frac{1}{R}\sum_{r=1}^Rv_{d,K,r}^2
\end{align*}
while an estimator for the between-imputation variance is given by
\begin{align*}
    B=\frac{1}{R-1}\sum_{r=1}^R\left(I_{n,M}^{OOB}(j,\mathcal{D}_n^r)-\widebar{I_{n,M}^{OOB}}(j)\right)^2.
\end{align*}
This leads to an estimator for the total variance given by
\begin{align*}
    v_{d,K,\text{rubin}}^2= \widebar{v_{d,K}^2}+\left(1+\frac{1}{m}\right)B.
\end{align*}
Following Rubin's rule \citep{Rubin1987,rubin2018multiple}, the resulting CIs for the population feature im-
portance scores are given by
\begin{align*}
    \left[I_{n,M}^{OOB}(j,\mathcal{D}_n)-z_{1-\alpha/2}\sqrt{v_{d,K,\text{rubin}}^2},~
    I_{n,M}^{OOB}(j,\mathcal{D}_n)+z_{1-\alpha/2}\sqrt{v_{d,K,\text{rubin}}^2}~\right].
\end{align*}
We will refer to them as 'Rubin-intervalls'. 
As stated above, this approach is heuristic as it is not backed by theoretical results. Nonetheless, we will later see in our numeric experiments that it leads to  significant improvements compared to the quick fixes by single imputations.

\section{Simulation Setting}
\label{Section 4}
To evaluate the performance of our six imputation strategies in conjunction with the Random Forest-based CIs for the population feature importance scores by \citet{ishwaran2019standard}, we adopt their data‐generating framework. This choice serves two purposes: first, it ensures comparability with prior work on tree‐based inference; second, it provides a diverse collection of nonlinear regression models to test the six imputation procedures. Below, we first describe the twelve regression models, including the form of regression function, feature space, error distributions, and sample sizes before we explain 
the mechanisms for inducing missingness and the operational details of our study.

\textbf{Regression Models}. We simulate $\mathbf{x}=(x_1,\dots,x_p)$ and responses $y$ according to twelve distinct functions. The settings span smooth non-linear terms (e.g., trigonometric or exponentials), higher-order interactions, indicator treshholds, and pure noise models, representing a broad spectrum of realistic data‐generating processes:
\begin{itemize}
    \item \textbf{Function 1 (Sine-Model)}: \( \mathbf{x} \sim \mathcal{U}([0,1]^{n \times p}) \), \( y = 10\sin(\pi x_1 x_2) + 20(x_3 - 0.5)^2 + 10x_4 + 5x_5 + \epsilon \), \( \epsilon \sim N(0,1) \).
    \item \textbf{Function 2 (Radial-Distance Model)}: \( \mathbf{x} \) with \( x_1 \sim \mathcal{U}([0,100]) \), \( x_2 \sim \mathcal{U}([40\pi, 560\pi]) \), \( x_3 \sim \mathcal{U}([0,1]) \), \( x_4 \sim \mathcal{U}([1,11]) \), others \( \mathcal{U}([0,1]) \), \( y = \sqrt{x_1^2 + (x_2 x_3 - (x_2 x_4)^{-1})^2} + \epsilon \), \( \epsilon \sim N(0,125) \).
    \item \textbf{Function 3 (Arctan Model)}: \( \mathbf{x} \) as in Function 2, \( y = \arctan((x_2 x_3 - (x_2 x_4)^{-1})/x_1) + \epsilon \), \( \epsilon \sim N(0,0.1) \).
    \item \textbf{Function 4 (Mixed Polyonomial Model)}: \( \mathbf{x} \sim \mathcal{U}([-1,1]^{n \times p}) \), \( y = x_1 x_2 + x_3^2 + x_4 x_7 + x_8 x_{10} - x_6^2 + \epsilon \), \( \epsilon \sim N(0,0.1) \).
    \item \textbf{Function 5 (Treshhold Polyonomial-Exponential Model)}: \( \mathbf{x} \sim \mathcal{U}([-1,1]^{n \times p}) \), \( y = \mathbb{I}(x_1 > 0) + x_2^3 + \mathbb{I}(x_4 + x_6 - x_8 - x_9 > 1 + x_{10}) + e^{-x_2^2} + \epsilon \), \( \epsilon \sim N(0,0.1) \).
    \item \textbf{Function 6 (Mixed Polyonomial-Exponential Model)}: \( \mathbf{x} \sim \mathcal{U}([-1,1]^{n \times p}) \), \( y = x_1^2 + 3x_2^2 x_3 e^{-|x_4|} + x_6 - x_8 + \epsilon \), \( \epsilon \sim N(0,0.1) \).
    \item \textbf{Function 7 (Treshhold Model)}: \( \mathbf{x} \sim \mathcal{U}([-1,1]^{n \times p}) \), \( y = \mathbb{I}(x_1 + x_4^3 + x_9 + \sin(x_2 x_8) + \epsilon > 0.38) \), \( \epsilon \sim N(0,0.1) \).
    \item \textbf{Function 8 (Mixed Log-Exponential Model)}: \( \mathbf{x} \sim \mathcal{U}([0.5,1]^{n \times p}) \), \( y = \log(x_1 + x_2 x_3) - e^{x_4/x_5 - x_6} + \epsilon \), \( \epsilon \sim N(0,0.1) \).
    \item \textbf{Function 9 (Mixed Polynomial-Root-Floor model)}: \( \mathbf{x} \sim \mathcal{U}([-1,1]^{n \times p}) \), \( y = x_1 x_2^2 \sqrt{|x_3|} + \lfloor x_4 - x_5 x_6 \rfloor + \epsilon \), \( \epsilon \sim N(0,0.1) \).
    \item \textbf{Function 10 (Mixed Power-Root Model)}: \( \mathbf{x} \sim \mathcal{U}([-1,1]^{n \times p}) \), \( y = x_3 (x_1 + 1)^{|x_2|} - \sqrt{x_5^2 / (|x_4| + |x_5| + |x_6|)} + \epsilon \), \( \epsilon \sim N(0,0.1) \).
    \item \textbf{Function 11 (Mixed Trigonometric Model)}: \( \mathbf{x} \sim \mathcal{U}([-1,1]^{n \times p} )\), \( y = \cos(x_1 - x_2) + \arcsin(x_1 x_3) - \arctan(x_2 - x_3^2) + \epsilon \), \( \epsilon \sim N(0,0.1) \).
    \item \textbf{Function 12 (Noise Model)}: \( x_j \sim N(0,1) \), \( y = \epsilon \), \( \epsilon \sim N(0,1) \).
\end{itemize}
To mimic real‐world settings, the number of features is set to $p=20$ for each model, by simply adding noise features, independent of $y$. That is, if $y=f(x_1,\dots,x_q)+\epsilon$ with $q<p$, we simply add iid features $x_{q+1},\dots,x_p$.
We vary the sample size as follows $n\in\{100,250,500\}$. The choice $n=250$ matches the original setting from \cite{ishwaran2019standard}, whereas $n=100$ and $n=500$ allow us to assess the impact of smaller resp. larger sample sizes on CI coverage. 

\textbf{Missing Mechanisms.} For the introduction of missings, we apply two different missing mechanisms. Missing Completely at Random (MCAR) randomly inserts missing values across all features in the dataset, with no connection to the data itself, ensuring that the missingness is entirely by chance. 
For the Missing at Random (MAR) algorithm, the 50th percentile (median) of \( x^{(j)} \) is computed, and the subset of rows with values below this threshold is identified
\[
\mathcal{I}_\text{MAR} = \{ i \in \{1,\dots,n\} \mid x_i^{(j)}  < \text{median}(x^{(j)} ) \}.
\]
From this subset, a proportion determined by the desired missing rate is sampled uniformly at random. 
In the selected rows, missing values are introduced independently in all features \( x^{(k)}  \) where \( k \ne j \).
This aims to mimic a pattern often seen in real-world datasets \citep{enders2010applied}. 

Within each MCAR or MAR scenario, we vary the missing rates (10\%, 30\%, and 50\%). MCAR (Missing Completely at Random) implies that missingness is independent of observed or unobserved data, with a fixed number of rows per column randomly set to NA, proportional to the missing rate. MAR (Missing at Random) means missingness depends on observed data (e.g., values in the second column below its median determine missingness in other columns), with a fixed number of NAs per affected column, proportional to the missing rate and the number of selected rows, or a fixed number via MCAR fallback if no rows are selected.

\textbf{Conducting the Experiments.}
For each of the $12\cdot3\cdot2\cdot3=216$ combinations of Model (Functions 1--12), sample size ($\in\{100,250,500\}$), missing mechanisms (MAR or MCAR) and missing rates ($\in\{10\%, 30\%, 50\%\}$), we conduct $n_{sim}=1,000$ independent simulation runs. Within each run the following steps are carried out:\\
\begin{enumerate}
    \item 
\textbf{Data generation:}
     Generate a dataset with the specified sample size of $n=250$ and $p=20$
 features according to the chosen regression models, including noise features independent of the response, as described in \cref{Section 4}.
 \item \textbf{Introducing Missing Values:} Introduce missing values under the MCAR or MAR mechanism with the specified missing rates.
 \item \textbf{Imputation:}
Apply each of the three imputation methods: 
\begin{itemize}
    \item Multiple Imputation by Chained Equations (MICE, using the \texttt{mice} package with predictive mean matching, \texttt{m=5}, \texttt{maxit = 5}),
    \item Chained Random Forests (\texttt{missRanger} package, with 250 trees, a node size of 5 and \texttt{pmm.k=5}),
    \item XGBoost-based imputation (\texttt{mixgb} package, supporting multiple imputation).
\end{itemize}
\texttt{mice} and \texttt{mixgb} yield $m=5$ completed data sets; \texttt{missRanger} returns a single completed data set.
 \item \textbf{Compute RFPIM:} 
Fit a Random Forest via \texttt{rfsrc} \citep{ishwaran2025fast}
with \texttt{ntree = 250}, \texttt{nodesize = 5},
\texttt{samptype = "swr"}, and \texttt{block.size = 1}. Compute the RFPIM $I_{n,M}(j)$, for each feature $j$ in each complete dataset, as defined in \cref{Section 2}.
\item \textbf{CI Construction:} Construct 95\% CIs ($\alpha = 0.05$) for the true feature importances using the two approaches: Imputing the data by (multiple) imputation and getting 
a single complete dataset on which the
Delete-$d$-Jackknife is determined.
And as an alternative for the multiple Imputation methods, the Delete-$d$-Jackknife is detemined on the \texttt{m=5} complete datasets and then combined by Rubin's rules.
The computation of the Delete-$d$ Jackknife variance estimator is done with \texttt{subsample.rfsrc} 
and \texttt{extract.subsample} from \cite{ishwaran2025fast}. We choose \texttt{B=100} and \texttt{subratio=sqrt(n)/n} as \cite{ishwaran2019standard}.
\item \textbf{Ground Truth}:
 Following \cite{ishwaran2019standard},
 estimate the
population feature importance scores by generating 1,000
complete data sets of size $n=250$ (no missingness) and
averaging their OOB importances $I_{n,M}^{\text{OOB}}(j)$. For the noise features the true feature importance scores are set to 0.
\item \textbf{Evaluation:} 
For each replicate and each imputation approach, record whether the
corresponding CI covers the ground truth 
feature importance.
Coverage rates are summarised over the 1,000 replicates per experimental condition. 
 \end{enumerate}
All simulations and computations are conducted with \textbf{R} Version 4.4.0 \citep{R}.

\section{Simulation Results}
\label{Section 5}
We evaluate the CIs on two criteria: coverage and interval length. 
Following \cite{ishwaran2019standard},
we stratify the features by their 'true' feature importance score
to account for signal strength. In our experiment, there are 240 features (12 functions with $p=20$ features). Their ground truth feature importance scores induce three groups: 
\begin{itemize}
\item \textbf{Group 1 - noise (zero importance)}: 185 features with no effect on $y$ and true importance 0.
\item \textbf{Group 2 - high-importance}: the top 10\% of all features 
(24 features) by true importance.
\item \textbf{Group 3 - moderate-importance}: the remaining 31 non-noise features. This group corresponds to the 75th-90th percentile group in \cite{ishwaran2019standard}.
\end{itemize}


Throughout the paper we display results for MAR and discuss them in detail. Analogues plots for MCAR are given in the supplement and their results are only summarized here.


\subsection{CI Coverage under MAR} 
\begin{figure}[h!]
    \centering
    \includegraphics[width=\textwidth]{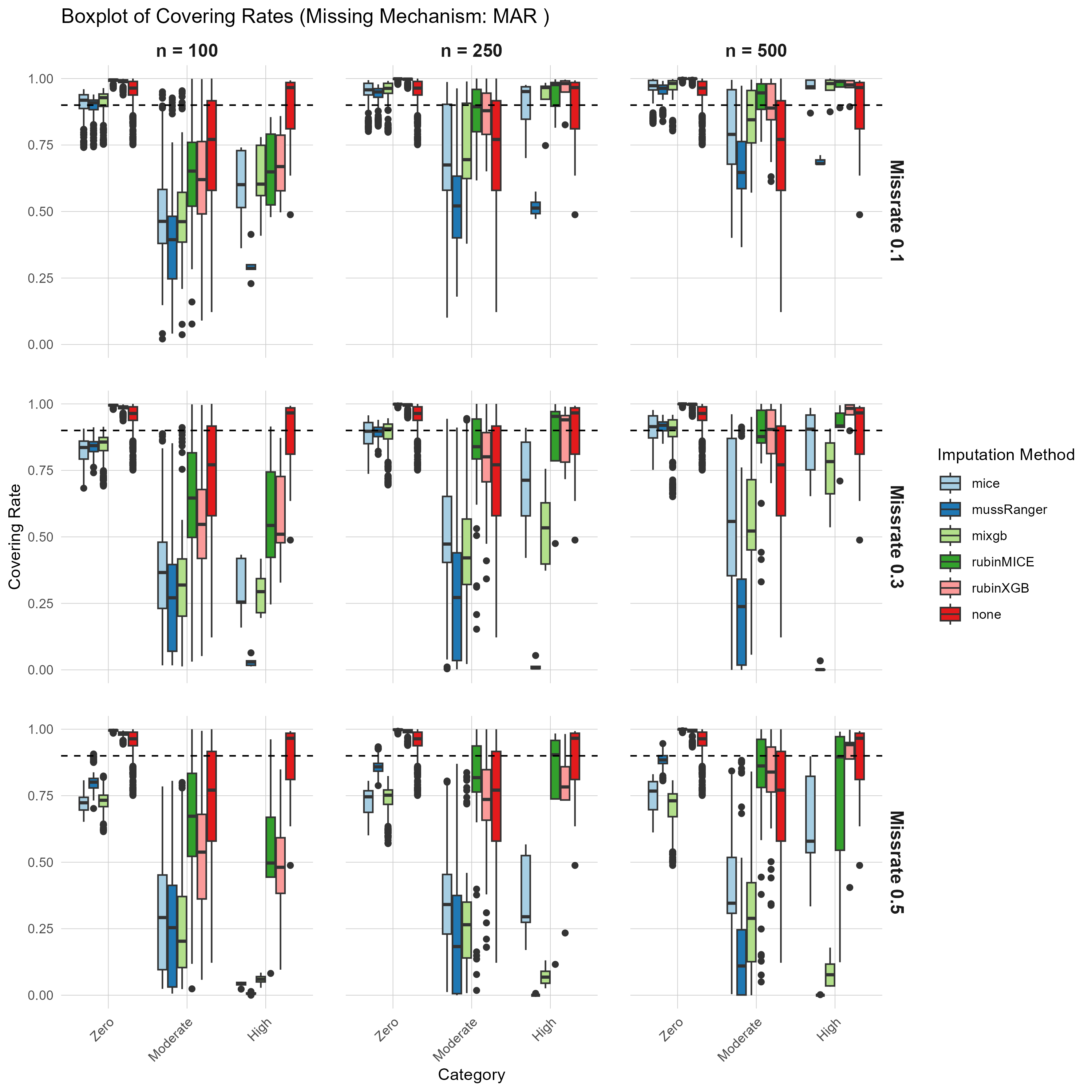}
    \caption{Boxplot-Matrix of coverage rates for MAR, stratified by Missrate (0.1, 0.3, 0.5), sample size (100, 250, 500) and category (Zero, Below 90\%, Above 90\%)}
    \label{fig:plot_matrix_mar}
\end{figure}

Figure~\ref{fig:plot_matrix_mar} shows boxplots of empirical coverage for each imputation method, stratified by feature group (noise, moderate, high), sample size ($n=100,250,500$ from left to right) and missing rate ($10\%, 30\%, 50\%$ from top to bottom). Within each subplot the methods are displayed from left to right in the following order: 1. single‐imputation variants (\texttt{mice}, missRanger, mixgb; all with PMM=5), 2. multiple imputation variants with Rubin's rule (\texttt{mice} and \texttt{mixgb}, each with $R=5$ imputations), and 3. the results without missings ('none'; for $n=250$), that serve as benchmark and are equal for all subplots.


\textbf{Group 1 (zero importance).}
CIs are quite conservative and attain or exceed, the nominal 95\% level at a missing rate of 10\% for all studied sample sizes. This also confirms the results from \cite{ishwaran2019standard}. 
For growing missing rates, only the Rubin-intervals can conserve this behavior. In particular, at 50\% missing rate, the coverage of the Ishwaran intervals on the imputed data fall below 90\%.

\textbf{Group 3 (moderate importance).}
For $n=100$, coverage collapses for all methods - even at 10\% missing rate. Therefore CIs 
for Feature Importance based on Random Forest are not reliable at this sample size. For the larger sample sizes ($n=250$ and $n=500$) the Rubin-Intervals are superior
in comparison to the Ishwaran intervals on imputed data.
 Especially if the imputation is made with
\texttt{missranger} we find extremely low coverage rates of under 50\%, which decrease even further for growing missing rates.
The Rubin intervals, on the other hand have even larger coverage rates than the intervals without imputation but they pay for this robustness with increased CI length (\cref{fig:plot_matrix_mar}).

\textbf{Group 2 (high importance).} 
In comparison to Group 3, CIs for high-effect features have larger coverage rates except for the scenario with 50\% missing rate. In this case \texttt{missranger} and \texttt{mixgb}
produce worse coverages.
The effect of missrate and $n$ is similar as for the moderate effect features. At 30\% missing rate and $n\geq 250$ only the Rubin intervals maintain coverages of at least $90\%$.

\noindent 
\textbf{Interim conclusion:} 
Applying the adaptation of Rubin’s rules to multiply imputed data can prevent the worst as they consistently outperform single-imputation CIs. 
The difference between the two Rubin methods is narrow. Moreover, neither 
\texttt{mice} nor \texttt{mixgb} seem to have an advantage regarding coverage.

\subsection{CI Lengths under MAR} 
To compare interval lengths across different experimental conditions, the CI lengths are standardized relative to a reference interval from the complete dataset (see below).
For each experimental condition, 
the mean CI length is calculated. The standardized CI length for each observation is obtained by dividing the length of the interval 
by the mean length of the corresponding reference interval. This normalization ensures that the standardized CI lengths are dimensionless and expressed as ratios relative to the baseline condition without missingness, allowing for a consistent comparison across different scales of features. 

\begin{figure}[h!]
    \centering
    \includegraphics[width=\textwidth]{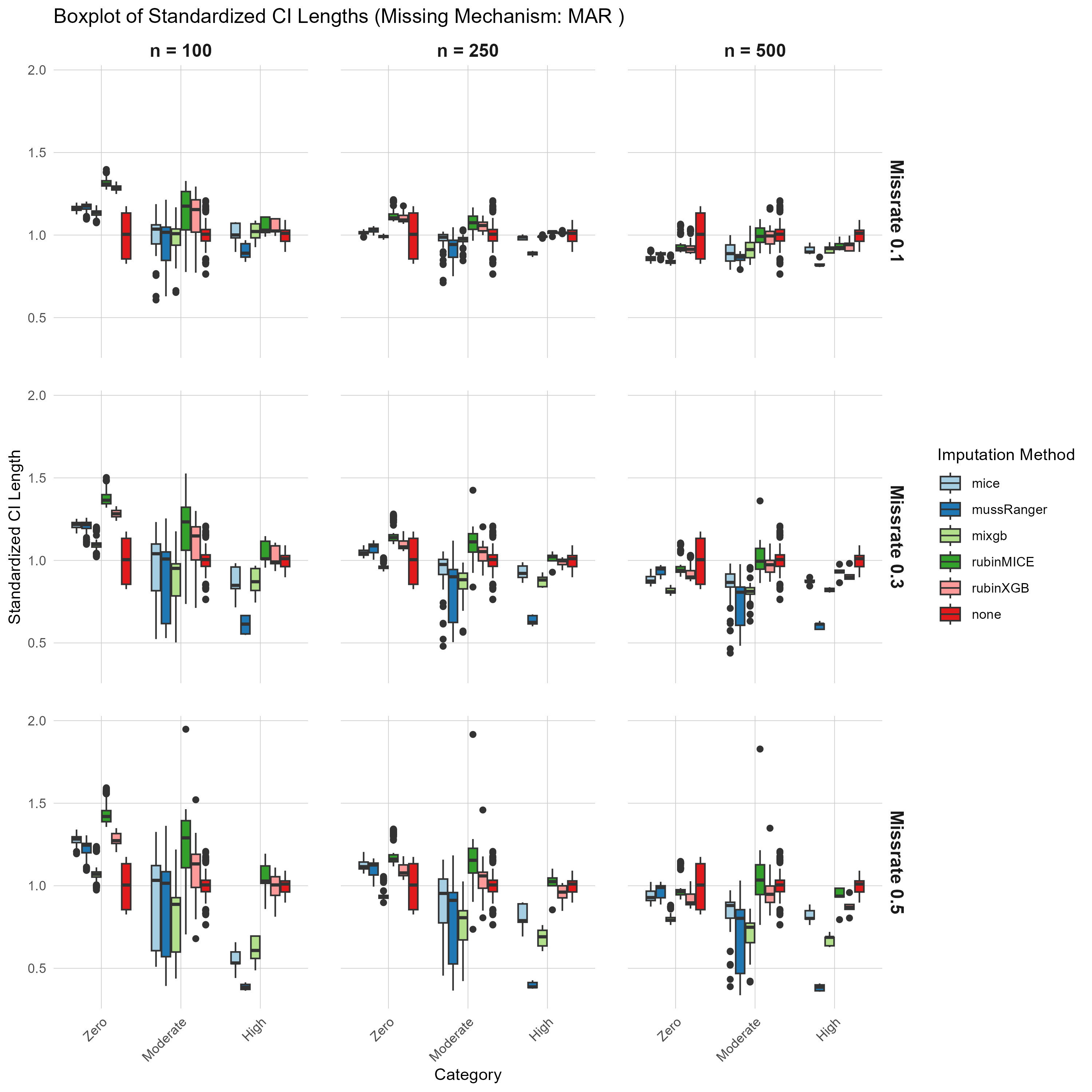}
    \caption{Boxplot-Matrix of standardized CI lengths for MAR, stratified by Missrate (0.1, 0.3, 0.5), sample size (100, 250, 500) and category (Zero, Below 90\%, Above 90\%)}
    \label{fig:plot_matrix_ci_length_mar}
\end{figure}

The results are again summarized in boxplots as above (Figure \ref{fig:plot_matrix_ci_length_mar}) and discussed below.

\textbf{Group 1 (zero importance).}
We observe very low variance in the CI lengths  with missings/imputation in comparison to the complete data intervals. The Rubin-intervalls are the largest, with the \texttt{mixgb} version being a little shorter than the \texttt{mice} version. The difference in lengths gets smaller with growing $n$. Moreover, the median length of all imputation methods is closer to the median of the CIs constructed on the complete data, when $n$ is larger.

\textbf{Group 3 (moderate importance).}
The behaviour mirrors the one from the noise group. 
The Rubin-intervals are still the 
longest, but with an increased gap between the \texttt{mixgb} and the \texttt{mice} version. For 30\% and 50\% missing rate and
$n=500$ the \texttt{mixgb} Rubin-Interval
seems to be the only interval whose box is shifted upwards. 
Another difference of moderate effect features and noise features is the
dispersion of the length of the complete data interval. For 10\% missing rate the dispersions of the six intervalsdon't show notable differences. With larger missing rates, however, the dispersion of lengths increases sharply for \texttt{mice}- and \texttt{missRanger}.

\textbf{Group 2 (high importance).}
The Rubin CIs again lead to the longest  intervals. 
\texttt{missRanger} has by far the shortest interval for high effect features across all settings. For $n=500$ all intervals based on imputation datasets are shorter than the intervals on the complete dataset. 

\noindent\textbf{Conclusion for MAR:} 
The Rubin-intervals are wide but safely conservative. Single imputation CIs tend to be too narrow, which likely explains their poor coverage.


\subsection{CI Coverage and Length under MCAR}
The results under the MCAR framework are given in Figures~\ref{fig:plot_matrix_mcar} (coverage) and \ref{fig:plot_matrix_ci_length_mcar} (lengths) in the supplement. 
Qualitative conclusions mostly match MAR and some key takeaways are discussed below:

\textbf{Sample Size.} With $n=100$ all methods - including the complete data benchmark - exhibit undercoverage. RFPIM inference is therefore not recommended at such small samples. With growing sample sizes ($n=250, 500$), however, the coverages improve.

\textbf{Missing Rate.} As expected, coverage decreases as the missing rate increases. 
For the approaches without Rubin’s rule, coverage can even fall below 50\% when the missing rate is 30\% or 50\%.
If $n=500$, the Rubin Interval with \texttt{mice} delivers the most robust coverage. However, for 50\% missing rate it also struggles and none of the considered approaches have reliable coverages.

\textbf{Feature Type.}
For the noise features, the Rubin intervals are as conservative as the intervals from the complete data. For the features with an effect,
the moderate effect features are hardest to cover as CIs become more liberal. 
Whether this stems from the effect size per se or the specific choice of functions. 

\textbf{Imputation Approach.}
Under MAR, both Rubin aproaches (\texttt{mice} and \texttt{mixgb})
had similar coverage; under MCAR,  Rubin with \texttt{mice} is superior to Rubin with \texttt{mixgb}. Nevertheless, both lead to much better coverage  than the Ishwaran intervals on single imputed data. Thereby, imputation with \texttt{missRanger} performs worst with coverage rates close to zero for larger missing rates.

\textbf{Conclusion for MCAR:} Again, Rubin-intervals were the most promising approach. However, they only exhibit reliable coverages for the lowes missing rate of 10\%, particularly in combination with \texttt{mice}. For larger missing rate, no method lead to reliable coverages, while Rubin with \texttt{mice} stayed more robust.\\

 \textbf{Summary:} Overall, these results demonstrate that  
\cite{ishwaran2019standard} CIs for permutation importance scores 
are not suitable for imputed data and small samples. In fact, with sample size of $n=100$, these CIs shouldn't be used at all. For moderate missing rate (30\% or 10\%) and adequate sample size ($n$ at least $250$), the combination of \cite{ishwaran2019standard} CIs with one of the two multiple imputation aprraoches via Rubin’s rules was more promising. 
Thereby, multiple imputation with \texttt{mice} was mostly preferable as it led to more reliable coverages in case of 10\% (both, for MAR and MCAR) and 30\% (only for MAR) missing rate. 
For the highest missing rate of 50\%, the results are not reliable.



\section{Benchmark Study}
\label{Section 6}
To evaluate how the competing CI approaches behave on real data, 
we analyse five publicly available regression datasets whose features are fully observed. Below, we briefly describe each dataset used in the experiments.

\begin{itemize}
    \item \textbf{Boston Housing}: This dataset contains information about housing prices in the Boston area, with 506 instances and 13 features, including the median house value (\texttt{MEDV}) as the target \citep{boston_housing}.
    \item \textbf{California Housing}: Comprising 20,640 instances and 8 features, this dataset provides data on housing prices in California, again with the median house value as the target \citep{california_housing}.
    \item \textbf{Concrete Compressive Strength}: This dataset includes 1,030 instances with 8 features related to concrete mixtures, aiming to predict the compressive strength of concrete \citep{concrete_compressive_strength}.
    \item \textbf{Wine Quality}: The red wine quality dataset contains 1,599 instances and 11 features, with the quality score (\texttt{class}) as the target \citep{wine_quality}.
    \item \textbf{Bike Sharing}: This dataset records bike rental counts with 17,379 instances and 14 features, targeting the total rental count (\texttt{count}) \citep{bike_sharing}.
\end{itemize}

\noindent \textbf{Experimental Framework:} For each dataset we delete $10\%$ of all entries MCAR, then impute the missing values using \texttt{mice}, \texttt{missRanger}, and \texttt{mixgb}.  
On every completed data set, we compute the Ishwaran CIs,
choosing the subsample size for the Delete-$d$ Jackknife estimator $b = \sqrt{n}$ to reduce runtime on the larger datasets. 
The Rubin intervals are also obtained for the two multiple imputation approaches 
\texttt{mice} and \texttt{mixgb}. This process is repeated 100 times.

Since there is no known ground truth, we cannot compute coverage rates. Hence, we compare the length and location of the CIs.  Because features live on different scales, each CI is standardized, such that
the interval without imputation has center 0 and mean.

\begin{figure}[ht]
    \centering
    \begin{subfigure}[b]{0.48\textwidth}
        \centering
        \includegraphics[width=\textwidth]{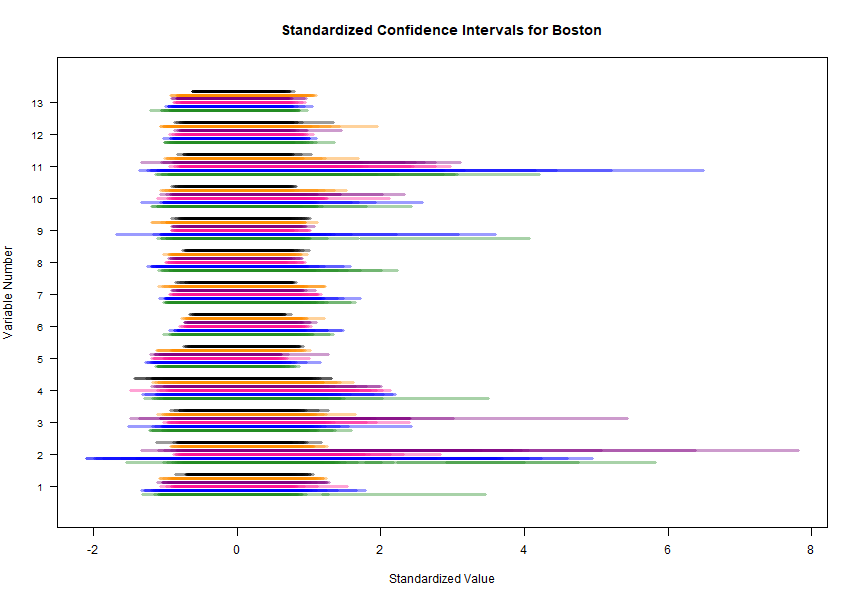}
        \caption{Boston}
        \label{fig:boston}
    \end{subfigure}
    \hfill
    \begin{subfigure}[b]{0.48\textwidth}
        \centering
        \includegraphics[width=\textwidth]{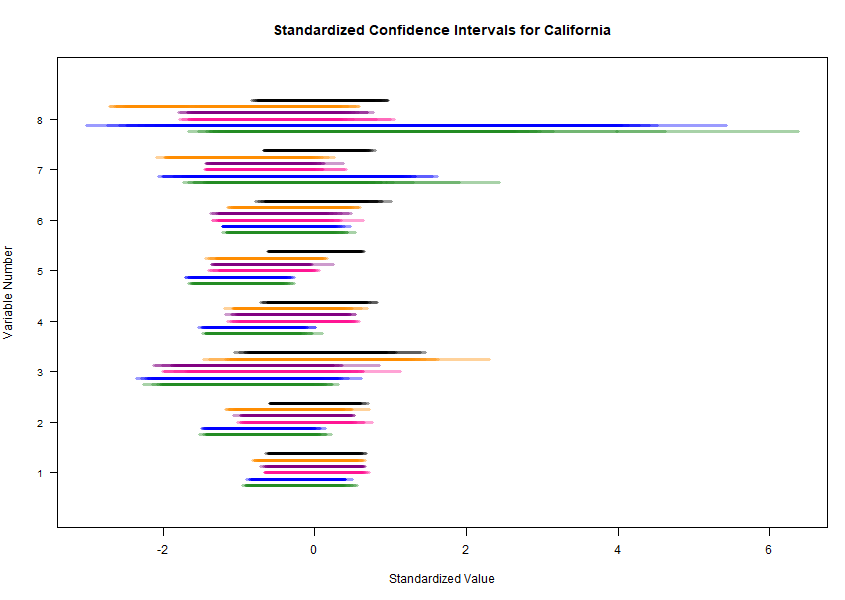}
        \caption{California}
        \label{fig:california}
    \end{subfigure}
    
    \vspace{10pt}
    \begin{subfigure}[b]{0.48\textwidth}
        \centering
        \includegraphics[width=\textwidth]{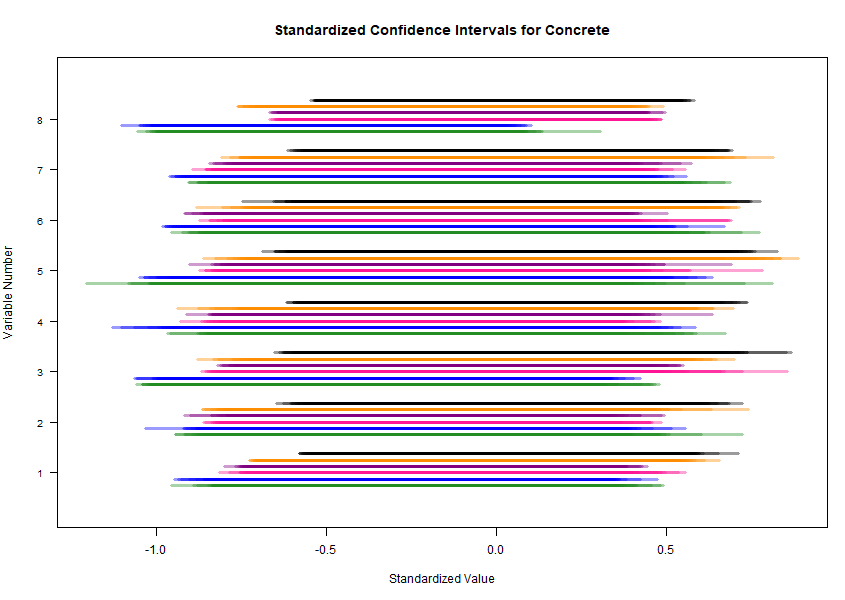}
        \caption{Concrete}
        \label{fig:concrete}
    \end{subfigure}
    \hfill
    \begin{subfigure}[b]{0.48\textwidth}
        \centering
         \includegraphics[width=\textwidth]{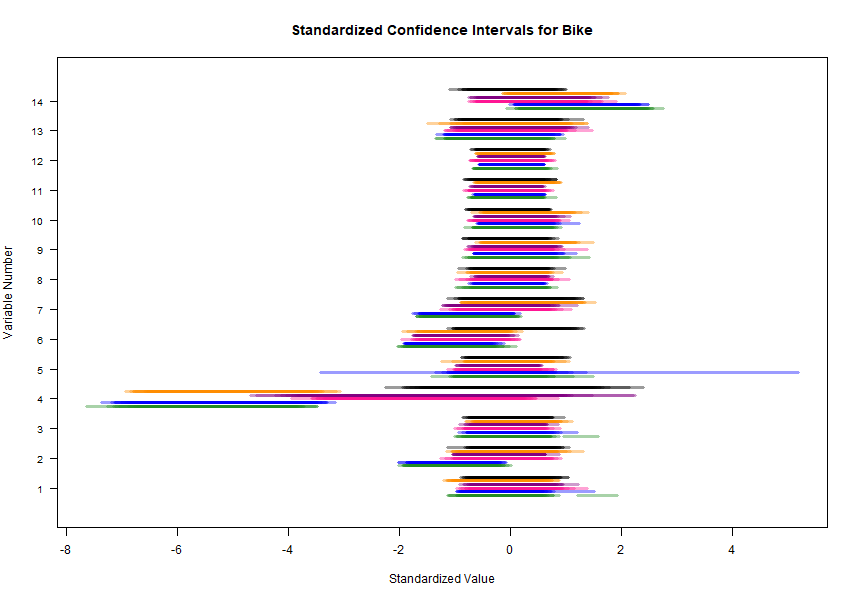}
        \caption{Bike}
        \label{fig:bike}
    \end{subfigure}
    
    \vspace{10pt}
    \begin{subfigure}[b]{0.57\textwidth}
        \centering
           \includegraphics[width=\textwidth]{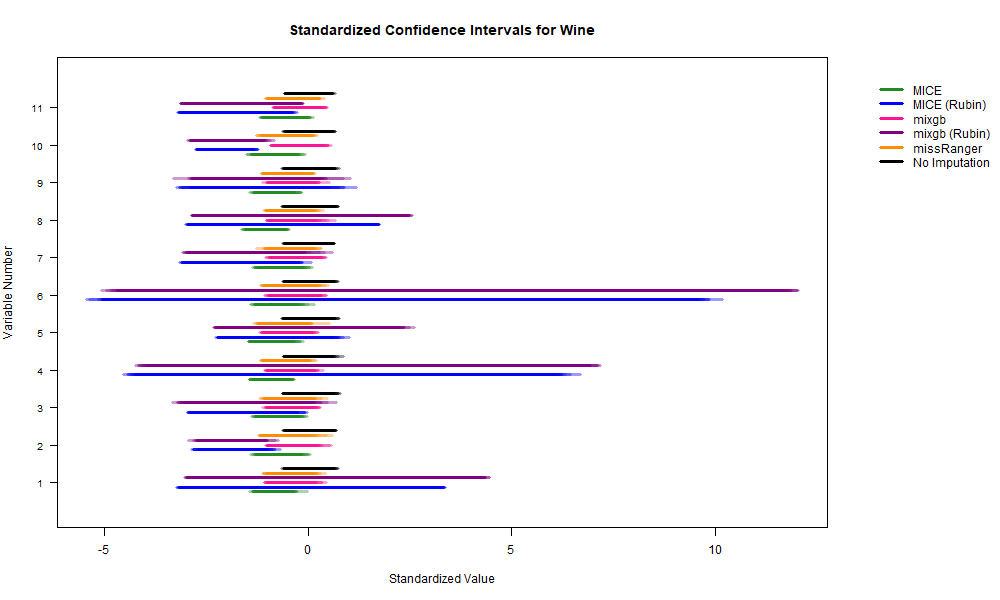}
        \caption{Wine}
        \label{fig:wine}
    \end{subfigure}
    
    \caption{Standardized CIs
    (mean length=1, center=0) relative to the complete data benchmark. Thereby x‑axes are scaled individually for
          each dataset. }
    \label{fig:all_datasets}
\end{figure}

\noindent \textbf{Results:} In Figure \ref{fig:all_datasets}, the standardized CIs for each feature across the five datasets are presented. 
The x-axes are scaled individually for each dataset, because of some outlier features.
Across all five datasets, the \texttt{mice}-based methods exhibit the largest shifts in the position of the CIs (i.e., changes in both lower and upper bounds) compared to the non-imputed dataset. In contrast, Random Forest imputation with \texttt{missRanger} results in smaller positional shifts. 

For the Boston dataset, CIs after \texttt{mice} imputation generally encompass those from the non-imputed dataset but have significantly wider upper bounds for some features. A similar pattern is observed for the Rubin-intervals with \texttt{mixgb}. In the California dataset, intervals from \texttt{mixgb} and \texttt{missRanger} show less pronounced positional shifts compared to the \texttt{mice} methods. The Concrete data set exhibit the smallest changes across all five datasets. In the Bike dataset, one feature exhibits significant shifts in the position of the CIs for both \texttt{missRanger} and \texttt{mice} methods, resulting in no overlap with the intervals from the non-imputed dataset. For this feature, only the two \texttt{mixgb} methods remain relatively robust, maintaining intervals with minimal positional shifts. The most notable results are observed for the Wine dataset. Here, \texttt{mice\_rubin} intervals are significantly shifted in their position compared to the non-imputed intervals and are, for some features, two to three times wider than those from other methods. In contrast, \texttt{missRanger} and \texttt{mixgb} methods show more moderate shifts and narrower intervals.

\noindent\textbf{Conclusions:} This benchmark study demonstrates that the performance of imputation methods is highly data-dependent. In fact, 
no single method uniformly minimises deviations from the complete-data CIs. 
Consequently, it does not provide a clear recommendation.

\section{Discussion}
\label{Section 7}
Missing data are ubiquitous in applied data analyses. This also holds for 
machine-learning applications. However, the properties of post-imputation inference remain poorly understood. In this study, we investigated the robustness of Ishwaran’s \citep{ishwaran2019standard} CIs for Random Forest Permutation Importance (RFPIM) in the presence of missing data. Through an extensive simulation study, we evaluated the performance of these intervals under various missing data scenarios (MCAR and MAR) and sample sizes ($n = 100, 250, 500$), comparing three different imputation methods: MICE, Random Forest-based imputation (\texttt{missRanger}), and XGBoost-based imputation (\texttt{mixgb}). 
Additionally, we proposed a heuristic approach based on Rubin’s rules to combine multiple imputations, aiming at improving coverage. An additional benchmark study on five real-world datasets complemented our simulations, examining the behavior of the CIs in practical settings. 

Our \textbf{key findings} from the simulation study were as follows: 1. Direct application of 
Ishwaran’s \citep{ishwaran2019standard} CIs on single imputed datasets fail, yielding to severe under-coverage, especially for moderate to high missing rates or smaller sample sizes. 2. Our newly proposed Rubin-inspired approach helps to mitigate these issues. It substantially improves coverage, often keeping 
the nominal 95\% level when $n$ is not too small ($n\ge 250$) or missing rates not too large ($\le 30\%$). The price is wider intervals. 3. Imputation approaches matter. 
While single imputation with \texttt{missRanger} often lead to the shortest intervals with severe under-coverage; \texttt{mice} and \texttt{mixgb} behave more stably, particularly \texttt{mice} with Rubin. In comparison, the benchmark study did not lead to a clear recommendation as real-data behavior varied considerably.

\textbf{Implications for practitioners:} Our results can be seen as a warning that 
imputation can destroy the nominal properties and thus the
interpretability of RFPIM CIs. If imputation cannot be avoided, multiple imputation in combination with Rubin’s rules is recommended. In general, the CIs have to be treated interval estimates with caution, particularly for moderate-signal features or high missing rate.

\textbf{Limitations and future work:} The Rubin-inspired procedures are heuristic and beside our numerical results no theorem guarantees its validity for RFPIM. Extending central limit theorems such as \citet{foge2024central} to incorporate imputation
        uncertainty thus is an important theoretical task. 

Ishwaran's \citep{ishwaran2019standard} CIs themselves need theoretical investigations. In particular, the theoretical assumptions underlying the asymptotic normality of RFPIM, as established by \cite{foge2024central}, are restrictive (independent features, i.i.d. setting, fixed dimensions etc.) and may limit the generalizability of our findings. Relaxing these and proving
        consistency of the Delete-$d$-Jackknife variance estimator
        for $\mathbb{V}\text{ar}(I_{n,M}^{OOB}(j))$ remains unclear. Moreover, only one specific MAR mechanism was studied and other realistic scenarios as well as MNAR frameworks are of interest. 

\textbf{Summary:} Until more results, particularly theoretical, we recommend to  use the proposed multiple imputation with Rubin’s rule while nevertheless interpreting RFPIM CIs from imputed data with appropriate scepticism.
\newpage





\begin{acks}
 This work has been supported by the Research Center Trustworthy Data Science and Security (https://rc-trust.ai), one of the Research Alliance Centers within the https://uaruhr.de.\\
The first author was supported by the Deutsche Forschungsgemeinschaft (DFG, German Research Foundation) - 314838170, GRK 2297 MathCoRe. The simulations were done on the Linux HPC cluster at Technical University Dortmund (LiDO3), partially funded in the course of the Large-Scale Equipment Initiative by the German Research Foundation (DFG) as project 271512359.
\end{acks}


\textbf{Statement:}
 During the preparation of this work, the authors used Grok 3 in order to improve language. After using this tool, the authors reviewed and edited the content as needed and take full responsibility for the content of the published article.

\bibliography{bibliography}   

    \setcounter{algocf}{0}

\newpage
\appendix

\begin{figure}[h!]
    \centering
    \includegraphics[width=\textwidth]{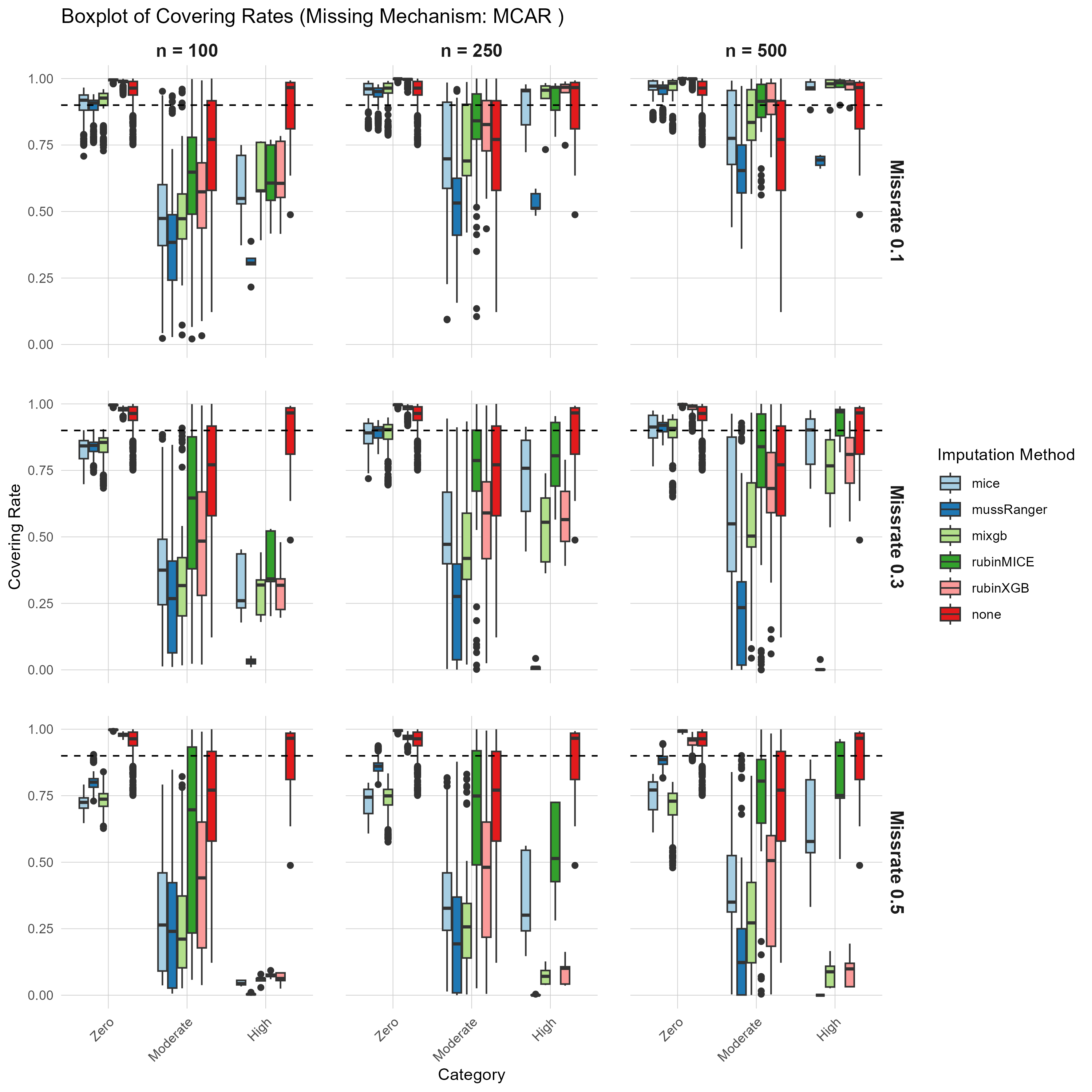}
    \caption{Boxplot-Matrix of coverage rates for MCAR, stratified by Missrate (0.1, 0.3, 0.5), sample size (100, 250, 500) and category (Zero, Below 90\%, Above 90\%)}
    \label{fig:plot_matrix_mcar}
\end{figure}

\newpage

\begin{figure}[h!]
    \centering
    \includegraphics[width=\textwidth]{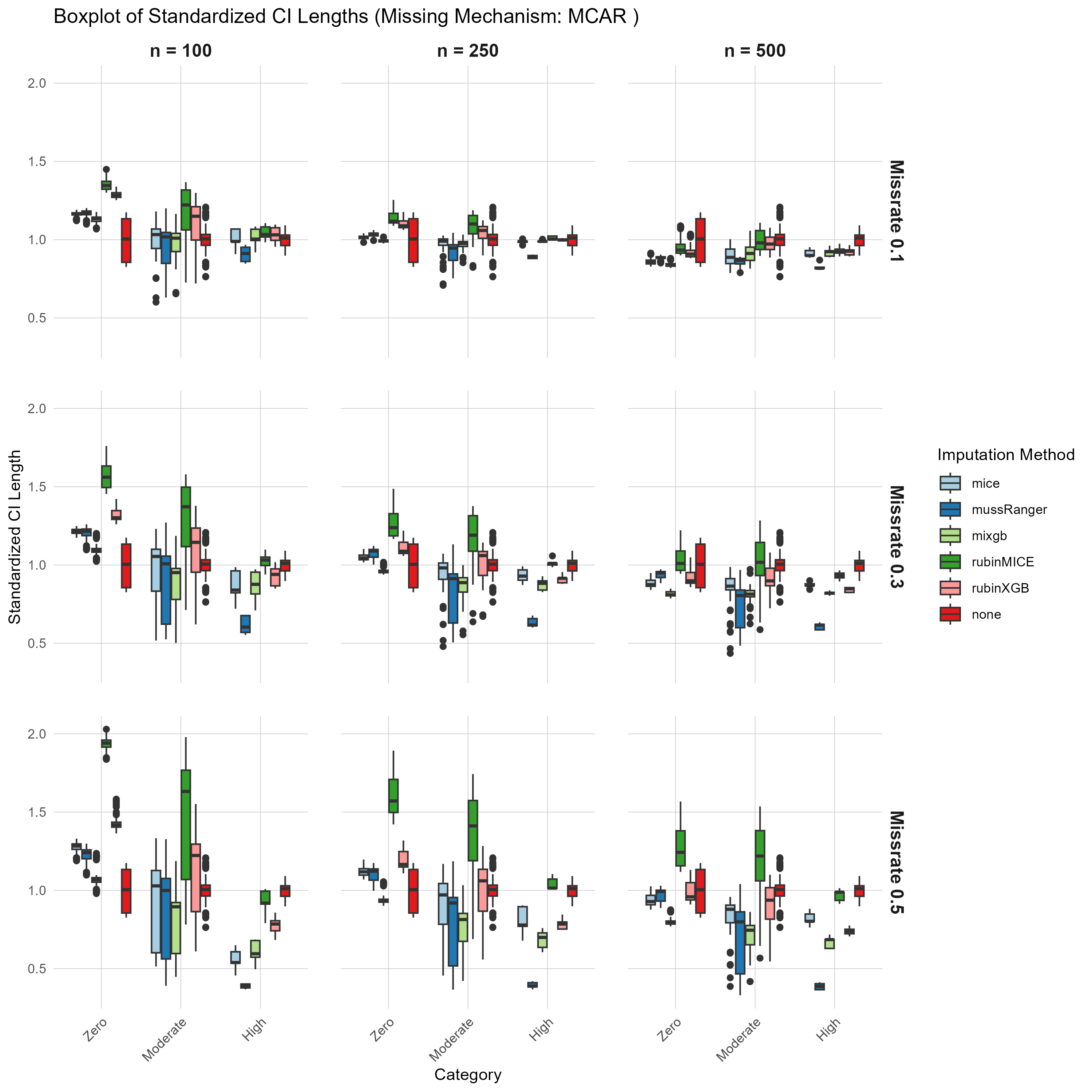}
    \caption{Boxplot-Matrix of standardized CI lengths for MCAR, stratified by Missrate (0.1, 0.3, 0.5), sample size (100, 250, 500) and category (Zero, Below 90\%, Above 90\%)}
    \label{fig:plot_matrix_ci_length_mcar}
\end{figure}

\end{document}